\documentclass[referee]{aa}
\usepackage{graphicx}
\begin{document}
\title{
Asteroseismology of exoplanets hosts stars : 
tests of internal metallicity 
}
\author{ M. Bazot\inst{1} \and S. Vauclair\inst{1}
}
\offprints{M. Bazot}
\institute{ Universit\'e Paul Sabatier, Observatoire Midi-Pyr\'en\'ees,
  CNRS/UMR5572,
 14 av. E. Belin, 31400 Toulouse, France
}
\titlerunning{ Asteroseismology of exoplanets hosts stars }
\authorrunning{ Bazot and Vauclair }
\date{Received; accepted}
\abstract{
Exoplanets host stars present a clear metallicity excess 
compared to stars without detected planets, 
with an average overabundance of 0.2 dex. This excess
may be primordial, in which case the stars should be overmetallic down
to their center, 
or it may be due to accretion in the
early phases of planetary formation, in which case the stars
would be overmetallic only in their outer layers. In the present paper,
we show the differences in the internal structure of stars, according to 
the chosen scenario. Namely two stars with the same observable parameters
(luminosity, effective temperature, outer chemical composition) are
completely different in their interiors according to their past histories, 
which we 
reconstitute through the computations of their evolutionary tracks. 
It may happen that stars with an initial overmetallicity 
have a convective core while the stars which suffered accretion
do not. We claim that asteroseismic studies of these exoplanets
host stars can give clues about their internal structures and metallicities,
which may help in understanding planetary formation. 
}

\maketitle

\section{Introduction} 

Since the historical discovery of the first exoplanet by Mayor 
et al \cite{mayor95}, 
more than one hundred and fifteen of these objects have presently been detected
(Mayor \cite{mayor03}, Santos et al \cite{santos04}).
Due to technical limitations, the systems which have been observed are quite 
different from the solar system. They usually consist in a solar-type star
surrounded by one or several Jupiter-like planets, orbiting
 at a short distance of the parent star.

 A striking observation about exoplanets host stars concerns their metallicities
 which, on the average, are larger than those of stars without detected planets,
 by 0.2 dex (e.g. Gonzalez \cite{gonzalez97}, Gonzalez \cite{gonzalez98}, 
Santos et al. \cite{santos01}, \cite{santos03} and \cite{santos04}, Murray et al.
\cite{murray01}, Martell \& Laughlin \cite{martell02}). The origin of this overmetallicity 
is still a subject of debate and is 
 directly related to the process of planetary formation. 

 Two extreme scenarios have been proposed to account for this observed overmetallicity :
we discuss them separately although it is not excluded that they both occur
together in the same stars. 
The first one assumes that the stellar systems with planets were formed out of 
an already metal-rich protostellar gas (see Santos et al. \cite{santos04} for 
a recent review on the subject). The subjacent 
idea is that a larger metallicity helps forming planets as it implies a larger 
density of grains and more frequent collisions among heavy particles. As a consequence it 
becomes easier for planetesimal to form and accrete matter (e.g. Pollack et al. 
\cite{pollack96}) 

The second scenario assumes no difference of metallicity in the protostellar 
gas. In this case the metallicity excess would be due to the accretion of newly 
formed planets in the early phases : due to their interaction with the disk, the
 planets may migrate towards the star and eventually fall on it. Evidences of
 planet migration come from the mere observation of giant planets close to the 
central stars : the formation scenarios assume that they have been formed further
 away and have spiralled down due to their interaction with the disk (e.g. Trilling et al
\cite{trilling98}).
 The basic difficulty is to understand the reason why the migration stops : it must
 be due to the absence of disk matter close to the star, which may be related
 to different effects. Within this
 framework, the accretion of some planets in the early epochs seems quite plausible. 
It may also be supported by the presence of $^6$Li in parent stars (Israelian et al
 \cite{israelian01}, \cite{israelian04}). 
As they are mostly composed of heavy elements, the 
accretion of these planets should increase the metallicity of the star in its
 outer layers (Murray et al \cite{murray01}). However the number of planets 
needed to explain an overmetallicity by 0.2 dex is large : a $3M_{\oplus}$
 terrestrial planet which could explain the $^6$Li/$^7$Li ratio in HD82943 
would increase the metallic abundance of the star by only $\simeq 17\%$ 
(Israelian et al \cite{israelian04}). Of course this depends on the chemical 
composition of the planets, but an increase of the metallicity as observed would
 need that the stars swallow several tens of earths, which is not completely
 excluded but seems very high.

A strong argument against the accretion scenario was raised by Santos et al \cite{santos01}, 
who pointed out that similar accretion rates on main sequence stars should lead
to a metal excess larger for stars with larger masses, as their convective zones 
are shallower (e.g. Pinsonneault et al \cite{pinsonneault01}). 
On the contrary, the observations show similar overmetallicities 
for stars of various effective temperatures. However this argument did not take 
into account double-diffusive 
convection, which should partially mix the accreted matter below the convective 
zone (Vauclair \cite{vauclair04}). Also accretion could occur before the zero
 age main sequence, in which case the convective zones are deeper.

Another argument has been used in favour of the primordial scenario : the existence 
of a correlation between the frequency of planets and the metallicity ; the more
 metal rich is the star, the larger is the probability of finding planets in orbit
 around it (e.g. Santos et al \cite{santos04}). This argument could however be used
 as a support of the accretion scenario as well : if many planets form around a star, 
the number of accreted ones is larger and the metallicity higher ; meanwhile the
 probability for the existence of an observable planet is also larger in this case. 

In the present paper, we suggest to use asteroseismology of exoplanets hosts stars 
to check whether they are overmetallic down to their center (``overmetallic models")
 or only in their outer layers 
(``accretion models"), which may help chosing between the two scenarios and lead 
to a better understanding of planetary formation.

Helioseismology has proved to be a powerful tool for solar physics, even for the 
detection of very small variations in the internal chemical composition (e.g.
 Bahcall et al \cite{bahcall92}, Christensen-Dalsgaard et al \cite{jcd96},
 Richard et al. \cite{richard96}, \cite{richard04}, Brun et al \cite{brun98},
 Turcotte et al \cite{turcotte98}). 
Asteroseismology is now taking its turn and is expected to lead to major advances in stellar 
physics. The technics are different for the stars and for the Sun as the stars are 
always seen globally while the solar surface can be locally analysed. Contrary 
to the Sun, only modes with low $l$-degree values may be detected in stars (from 
0 to 3 with the present asteroseismic technics). Even with such 
low degree modes, important clues can be obtained for the internal structure of the stars.
 Ground-based and space instruments (ELODIE, HARPS, MOST) are able to 
observe oscillations of solar-type stars. In the future, more precise
observations are still expected from new missions like CoROT and possibly 
Eddington. 

A good example
of such asteroseismic studies is the detection of acoustic modes 
in $\alpha$Cen A (Bouchy \& Carrier \cite{bouchy01}) using the ELODIE spectrograph in Haute 
Provence Observatory. Twenty eight modes of low degrees, ($l=0,1,2$) and radial 
orders between 15 and 25 were identified (Bouchy \& Carrier \cite{bouchy02}). These 
oscillation frequencies were used to provide an accurate description of the 
internal physics of the star (Thevenin et al \cite{thevenin02}, Thoul et al. \cite{thoul03}, 
Eggenberger et al. \cite{eggenberger04}).

In the present paper, we compute stellar models for the overmetallic and 
accretion cases, which are iterated so that they converge on the same ``observable" 
parameters (here we use $L, T_{eff}$ and chemical composition). The stars could not be
differentiated by classical observation, but we will show that they are quite 
different in their internal structure and that the differences can be
tested with asteroseismology. 

The computations and computational
 parameters of the models are described in section 2. Two couples of
models which are similar in their external parameters but with different
histories and internal structures are defined in section 3 : their 
differences are analysed in details. Section 4 is devoted to the
asteroseismic signatures of these differences. The results are 
summarized and discussed in section 5.

\section{The computations}

\subsection{The physical parameters}

The stellar models were computed with the Toulouse-Geneva code, including the most 
recent improvements in the physical inputs : new OPAL equation of state
 (Rogers \& Nayfonov, \cite{rogers02}) ; NACRE nuclear reaction rates 
(Angulo et al., \cite{angulo99}) ; OPAL opacities (Iglesias \& Rogers \cite{iglesias96}).
 Microscopic diffusion was included in all the models as described in Richard et al \cite{richard04}. 
The convective zones were computed in the framework of the mixing length theory,
 with a mixing length parameter calibrated on the Sun and identical in all models :
 $\alpha = 1.895$. No overshoot neither mixing were introduced in the present paper :
their effects will be treated in a forthcoming paper.
The oscillation frequencies of the models were computed using the 
Brassard et al. adiabatic pulsation code (\cite{brassard92}).

Evolutionary tracks (including standard pre main-sequence) were computed for stars 
with initial solar abundances (see table 1) modified
 by accretion in its outer layers (accretion models) and for stars with initial 
metallic excess (overmetallic models)

\subsection{The accretion scenario}

The accretion scenario assumes that the star is overmetallic only in its outer layers. 
This could be due to planetary material falling onto the star. We do not discuss here
 detailed scenarios for the accretion of planets, planetesimals or planetary material 
through evaporation. We assume, for a first approach, that the accreted matter has no
 hydrogen neither helium, while the other elements are present in solar relative abundances (this
assumption can be tested by spectroscopy when applied to a real star).
 Also the accretion process is treated as instantaneous compared to the evolution time scale and 
is assumed to take place on the ZAMS.
Except for hydrogen and helium,
 we set the surface abundances at the time of zero age main sequence as
 $X^{a}_{i}=X_{i}(1+f_{a})$ where f$_{a}$ is a free parameter linked to the accreted mass.
 The hydrogen and helium mass fraction are then recomputed such
that their ratio remains unchanged.

The effect of accretion on a standard model with solar initial abundances is shown in Fig. 1.
 We can see that the models with accretion are cooler than those without accretion which have 
the same internal abundances. 

\begin{figure}
\begin{center}
\includegraphics[angle=0,totalheight=\columnwidth,width=\columnwidth]{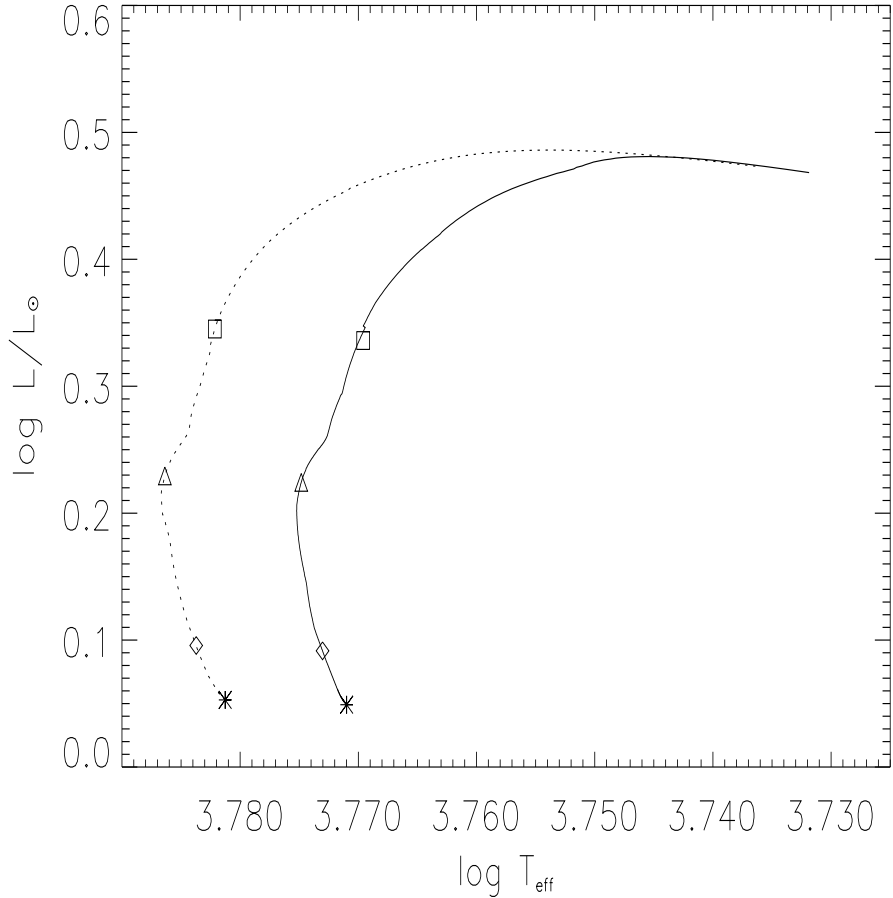}
\end{center}
\label{fig1}
\caption{Evolutionary tracks for main sequence models of 1.1 M$_{\odot}$ without accretion 
(dashed line) and with accretion (full line). The symbols on each track correspond to ages 
of 30 Myr (asterix), 1 Gyr (diamonds), 4 Gyr (triangles) and 6 Gyr (squares). Here the 
comparisons are given for two models with the same internal chemical composition. We can
 see how accretion moves the tracks towards smaller effective temperatures. In the following
 we will compare models with the same external chemical composition and different internal structure.}
\end{figure}

\subsection{The primordial scenario}

The primordial scenario assumes that the star and the planets were formed
out of initially overmetallic gas. This suggests that interstellar clouds inside the
 Galaxy may have metallic abundances varying by a factor of order two. We do not know 
however the helium abundance in such clouds. A possible assumption would be that it
follows the regression law found for extragalactic HII regions, namely :

\begin{equation}
Y=Y_{p} + Z\frac{\Delta Y}{\Delta Z}
\end{equation}
where Y$_{p}$ is the primordial helium ratio and $\frac{\Delta Y}{\Delta Z}$ the slope
 of the regression curve obtained from observations (Isotov \& Thuan \cite{isotov04}).

But the chemical history of intersellar clouds may be different from the 
overall chemical evolution of galaxies as the average stellar mass fraction may not be the same. 
In this case it is possible that helium remains solar while the metal abundances increase.

In the first case, the helium value for a metal increase by a factor 2 compared to the Sun
would be Y=0.33 while 
in the second case it would remain Y=0.27 as in the Sun. The two cases have been treated in our computations as two possible extreme values.

\subsection{Evolutionary tracks}

Figs. 2 and 3 present evolutionary tracks for accretion and overmetallic models, for a 
selection of stellar masses. In Fig. 2 the overmetallic models have a helium abundance 
Y=0.27 while in Fig. 3 the helium abundance is Y=0.33. The accretion models are the same 
in both figures. The effective temperature scale is different. In each figure we can see 
that the behavior of the evolutionary tracks is quite different for accretion and 
overmetallic models. The models which have a metal excess down to their center have 
a convective core while the accretion models do not. Also a comparison between the two 
figures show that the overmetallic stars which also present a helium excess are hotter 
than those with a solar helium abundance.

Thus, to obtain the same external parameters (L, T$_{eff}$ and metallicity), with the same 
Y$_{0}$ and $\alpha$, we expect that an overmetallic star will have a larger mass than a 
star which has suffered accretion. This fact has two major consequences which will be 
discussed later. First the internal structures of the stars are quite different and 
can hopefully be distinguished with asteroseismic studies ; second the determination 
of the masses of exoplanets may have to be reviewed if their host star is heavier than 
previously assumed.

\begin{figure}
\begin{center}
\includegraphics[angle=0,totalheight=\columnwidth,width=\columnwidth]{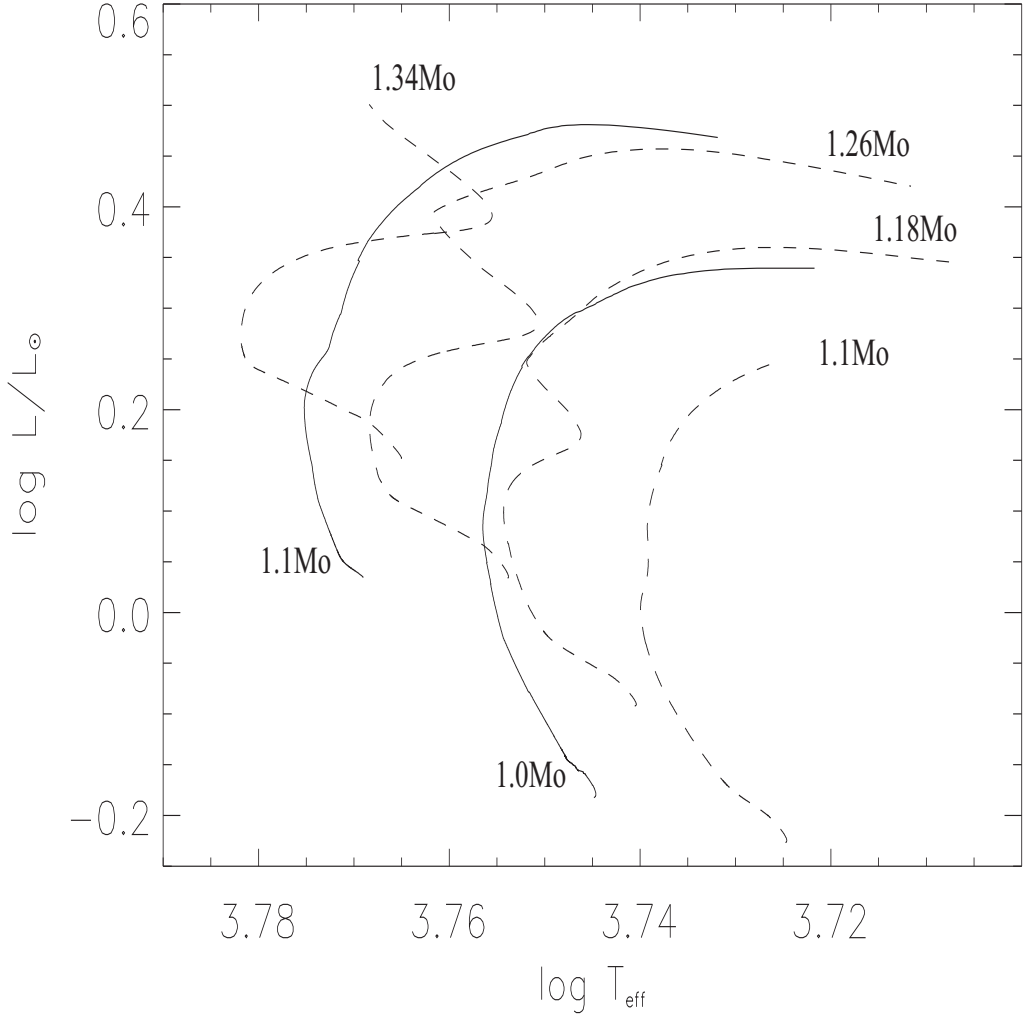}
\end{center}
\label{fig2}
\caption{Evolutionary tracks for overmetallic (dashed lines) and accretion 
(full lines) models. In these computations, the overmetallic models have initial 
values : Z=0.04 and Y=0.27 while the accretion models have initial values Z=0.02 
(increased to 0.04 in the outer layers at the beginning of the main sequence, due
 to accretion) and Y=0.27. For the overmetallic models, the tracks are plotted, from
 right to left, for the following masses : 1.1, 1.18, 1.26, and 1.34 M$_{\odot}$.
 For the accretion models, the tracks are plotted for 1.0 and 1.1 M$_{\odot}$.}
\end{figure}

\begin{figure}
\begin{center}
\includegraphics[angle=0,totalheight=\columnwidth,width=\columnwidth]{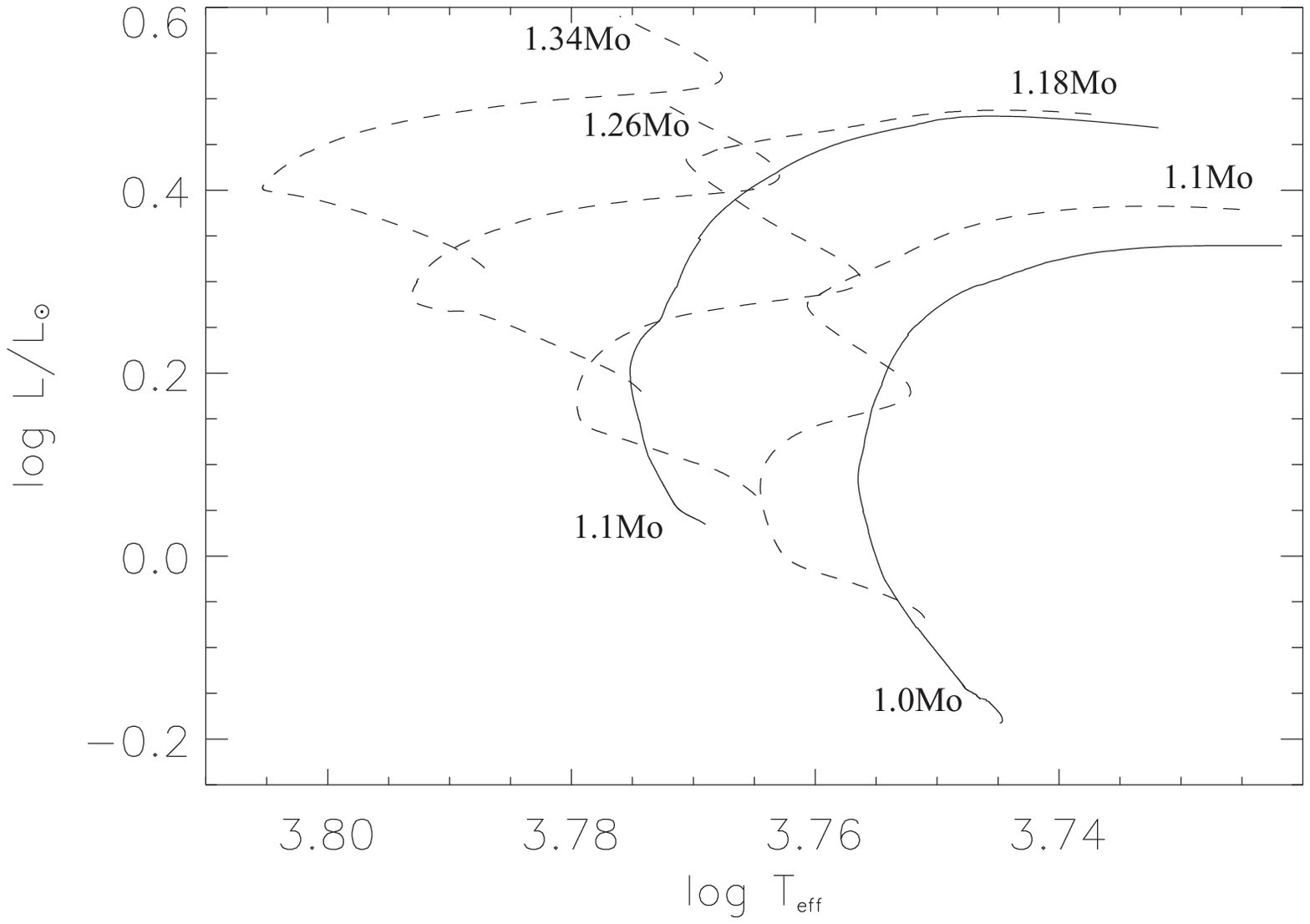}
\end{center}
\label{fig3}
\caption{Same as Fig. 2, except that here the overmetallic models have initial values : Z=0.04 and Y=0.33}.
\end{figure}

\section{Comparisons of overmetallic and accretion models with similar external parameters} 

\subsection{The models} 

We have computed evolutionary tracks for models with an initial overmetallicity and models
 with accretion. We chose to analyse two couples of models obtained with the same evolutionary
 track for the case of accretion, but two different tracks for the overmetallic case : one
 with an initial helium value $Y_0 = 0.27$ (they are called AC1 and OM1) and the second one
 with $Y_0 = 0.33$ (AC2 and OM2). The evolutionary tracks and the position of these models
 in the HR diagram are displayed on Fig. 4. As the aim of our computations was to compute
 models with similar external observable parameters, the evolutionary tracks correspond to
 different masses : $1.3 M_{\odot}$ and $1.15 M_{\odot}$ respectively for the overmetallic 
models, compared to $1.1 M_{\odot}$ for the pulluted ones.
A striking point is that a convective core develops in the overmetallic models, which gives
 the corresponding characteristic shape to the evolutionary track. On the other hand, the models
 with solar internal abundances which lie at the same position in the HR diagram do not develop
 any convective core. 

The masses of the models are given in Table 1 together with their ages, and their initial and final surface
 chemical compositions. Table 2 shows the effective temperatures, luminosities, radii,
 surface gravities, acoustic depths of the models (i.e. time needed for the acoustic
waves to travel along the stellar radius), as well as the geometrical and acoustic depths of the outer 
convective zones and the geometrical and acoustic radii of the convective cores if any. 
For an external observer, the stars are very similar inside each couple. The relative
 differences for T$_{eff}$ and $\frac{L}{L_{\odot}}$ are very small : respectively of
 order $10^{-5}$ and $10^{-4}$. The metallicity is more difficult to match exactly because
 of element diffusion along the evolutionary tracks : the relative difference is $7\%$ for
 models AC1 and OM1 and $4\%$ for models AC2 and OM2. In each case the differences between
 the models are not detectable through classical spectroscopic observations. Their internal 
structures are however very different, as seen below.

\begin{table}
\label{tab1}
\caption{Masses, ages and chemical composition of the stellar models. X$_0$, Y$_0$ and Z$_0$
 are the initial surface mass fractions of H, He and metals while X, Y and Z are the final ones.}
\begin{center}
\begin{tabular}{ccccccccc} \hline
\hline
Model & M$_{\star}$ & Age (Gyr) & X$_0$ & Y$_0$  & Z$_0$ & X & Y & Z \cr
 \hline
 AC1    & 1.1  & 5.578 & 0.7097& 0.2714 &  0.0189  & 0.7478 & 0.2137 & 0.00385   \cr 
 OM1    & 1.3  & 3.623 & 0.6844& 0.2714 &  0.0442  & 0.7221 & 0.2366 & 0.0413   \cr
 AC2    & 1.1   & 3.290 & 0.7097& 0.2714 &  0.0189  & 0.7287 & 0.2310 & 0.00403    \cr
 OM2    & 1.15  & 2.771 & 0.6260& 0.3297 &  0.0443  & 0.6589 & 0.2993 & 0.00418   \cr
 \hline
\end{tabular}
\end{center}
\end{table}

\begin{table*}[b]
\label{tab2}
\caption{Output parameters of the models : effective temperature, luminosity, radius, 
surface gravity, total acoustic depth, geometrical and acoustic depth of the outer
 convective zone, geometrical and acoustic radius of the convective core (if any).}
\begin{center}
\begin{tabular}{cccccccccc}
\hline
\hline
Model  & T$_{eff}$ (K) & $\frac{L}{L_{\odot}}$ & R$_{\star}$ (cm) 
& g (g.cm$^{-2}$)& t$_{\star}$ (s) & $\frac{R_{ec}}{R_{\star}}$ & $ \tau_{ec} $ & $\frac{R_{cc}}{R_{\star}}$ & $t_{cc}$ \cr
\hline
AC1    & 5902.36  & 2.029 &  9.50e10  & 16146  &  5300.6  & 0.7211   & 3407 &  -  &   -  \cr
OM1    & 5902.48  & 2.030 &  9.50e10  & 19067  &  4900.9  & 0.7397   & 3056 & 0.0671 & 124.4 \cr
AC2    & 5959.03  & 1.555 &  8.16e10  & 21881  &  4254.5  & 0.7369   & 2760 &  -  &  -  \cr
OM2    & 5959.36  & 1.555 &  8.16e10  & 22880  &  4162.6  & 0.7191   & 2566 & 0.0655 & 102.5 \cr
\hline
\end{tabular}
\end{center}
\end{table*}

\begin{figure}
\begin{center}
\includegraphics[angle=0,totalheight=\columnwidth,width=\columnwidth]{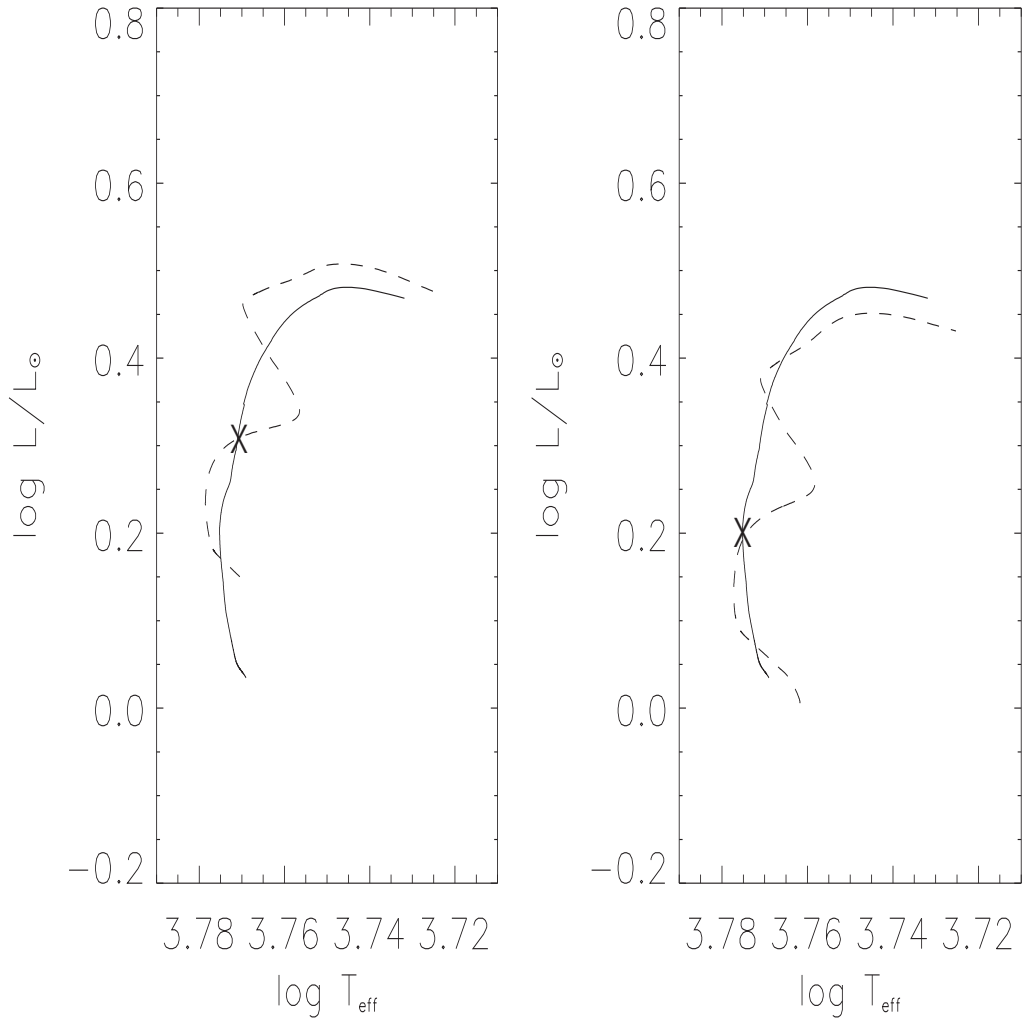}
\end{center}
\label{fig4}
\caption{ Evolutionary tracks for accretion models (full lines) and overmetallic models 
(dashed lines) ; in both graphs, the accretion models have a mass of 1.1 M$_{\odot}$ and 
a solar composition inside (see Table 1) ; in graph {\bf a.}, the overmetallic models
 have a mass of 1.3 M$_{\odot}$ and an initial helium mass fraction Y$_0$=0.27 (solar like) ;
 in graph {\bf b.}, the overmetallic models have a mass of 1.15 M$_{\odot}$ and an initial 
helium mass fraction Y$_0$=0.33. The crosses identify the models AC1 and OM1 ({\bf a.}), 
and AC2 and OM2 ({\bf b.}) }
\end{figure}

\subsection{Internal structure and chemical composition} 

For each couple, the overmetallic model has a convective core, contrary to the accretion model.
 Figures 5 and 6 display the relative differences of the internal density, pressure, temperature and
 square sound velocity in the two couples of models as a function of the fractional radius. For a
 given variable {\it x}, the adopted notation is $\frac{\Delta x}{x} = \frac{x_{OM}-x_{AC}}{x_{AC}}$ 
where OM and AC stand respectively for ``overmetallic" and ``accretion". 
We can see that the differences are large, and the positions of the convective cores in the
 overmetallic models are clearly seen in the sound velocity graph.

Figures 7 and 8 display the internal chemical composition of the models. Figure 7 presents the abundance
 profiles of H, He, C, N, O, Ne, Mg and the heavier metals gathered in label Z for models AC1 and OM1.
 Fig. 8 presents the same graphs for models AC2 and OM2. The differences between the overmetallic
 and accretion models are clearly visible on these graphs.

\begin{figure}
\begin{center}
\includegraphics[angle=0,totalheight=\columnwidth,width=\columnwidth]{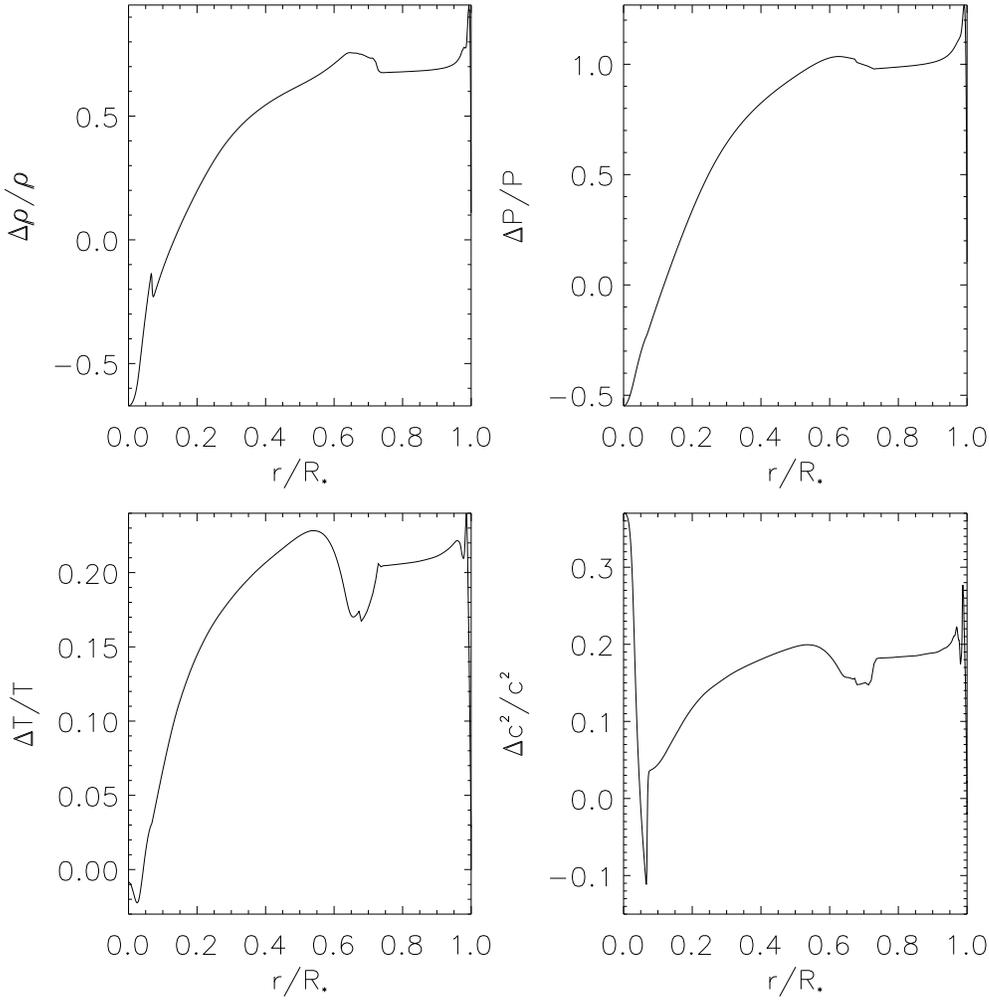}
\end{center}
\label{fig5}
\caption{Relative differences in density, pression, temperature and square of the sound speed
 between models AC1 and OM1. For any variable {\it x}, the ratio $\frac{\Delta x}{x}$ is
 defined as $\frac{x_{OM1}-x_{AC1}}{x_{AC1}}$.}
\end{figure}
 
\begin{figure}
\begin{center}
\includegraphics[angle=0,totalheight=\columnwidth,width=\columnwidth]{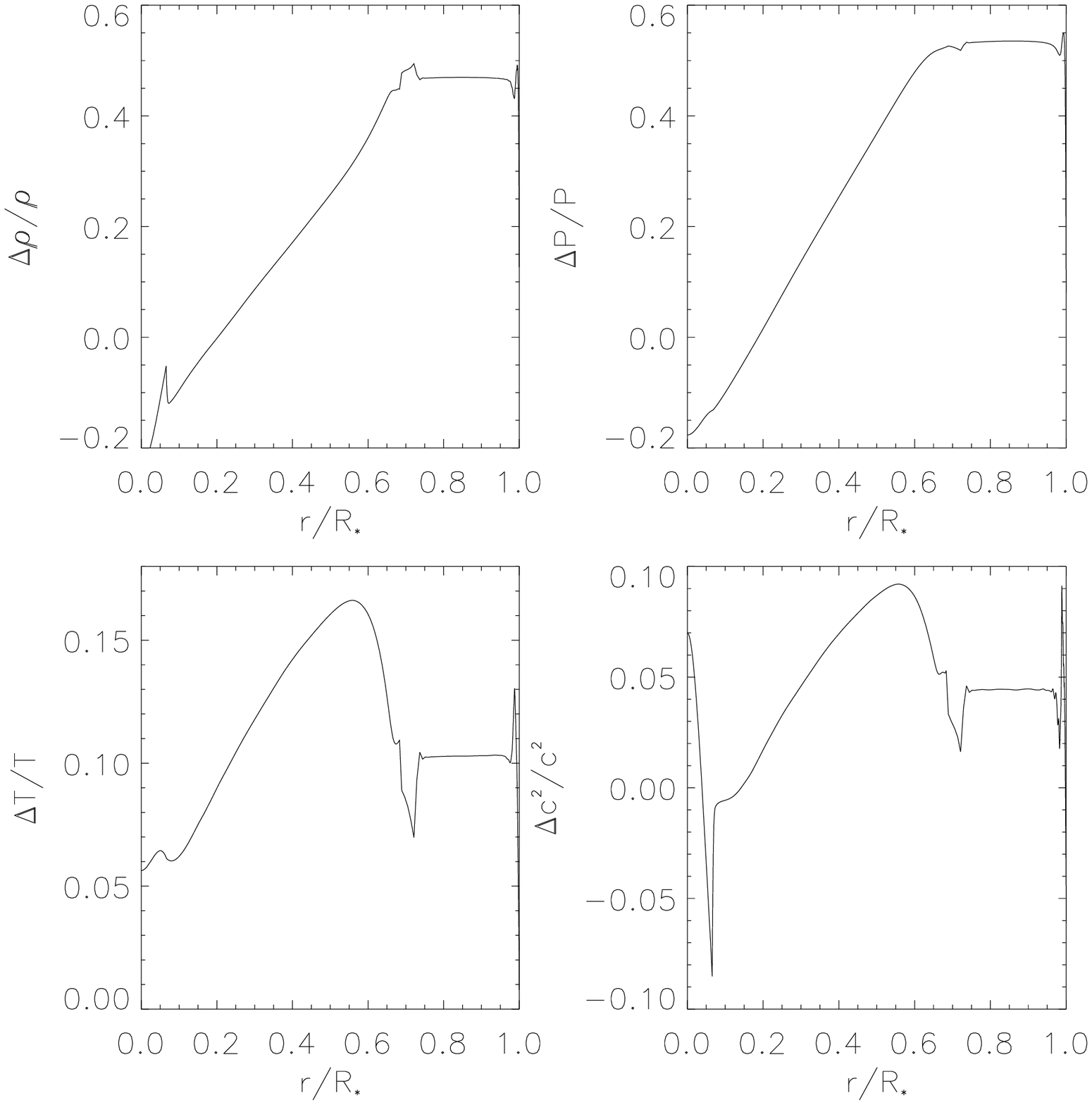}
\end{center}
\label{fig6}
\caption{Same as Fig.5, for models AC2 and OM2.}
\end{figure} 

\begin{figure}
\begin{center}
\includegraphics[angle=0,totalheight=\columnwidth,width=\columnwidth]{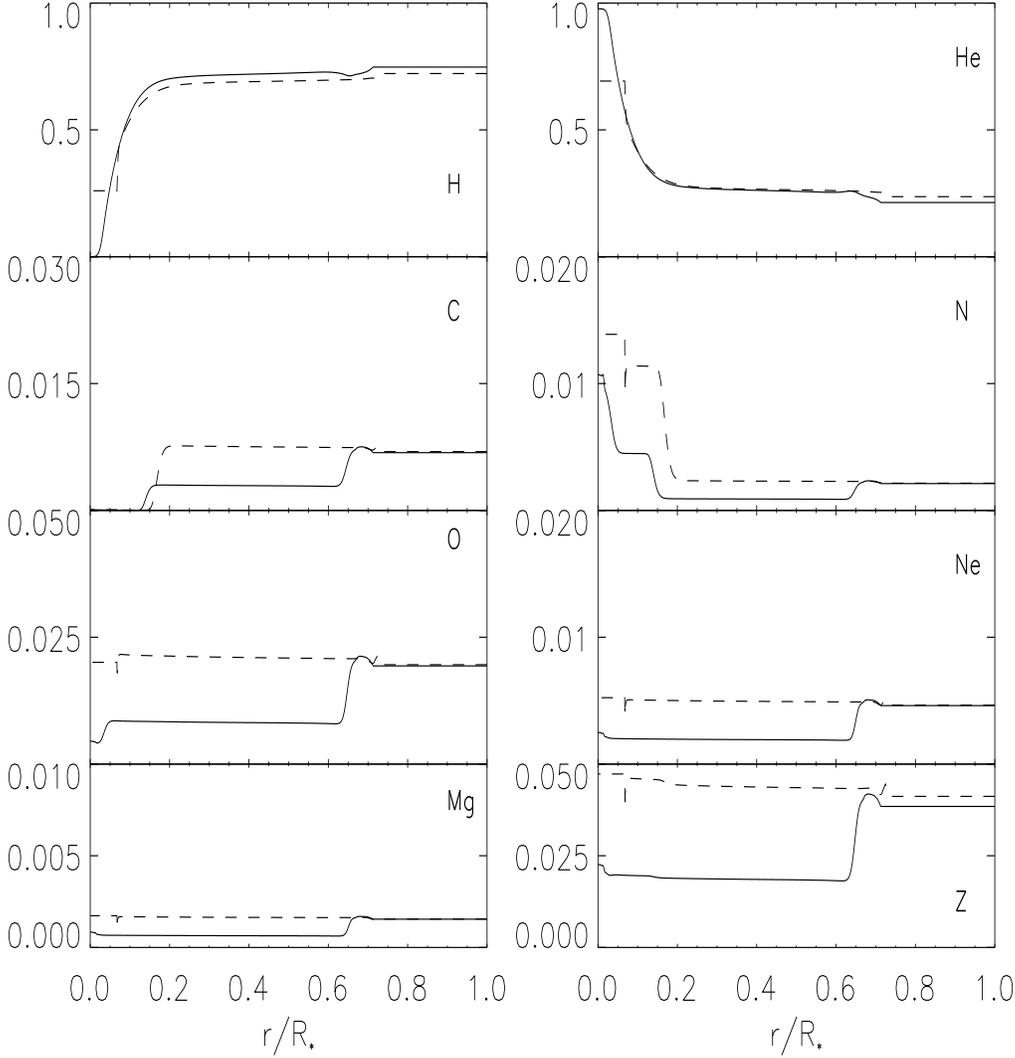}
\end{center}
\label{fig7}
\caption{Mass fractions of H, He, C, N, O, Ne, Mg and Z(which includes all the heavier elements), 
as a function of the fractional radius, in models AC1 (full lines) and OM1 (dashed lines).}
\end{figure}

\begin{figure}
\begin{center}
\includegraphics[angle=0,totalheight=\columnwidth,width=\columnwidth]{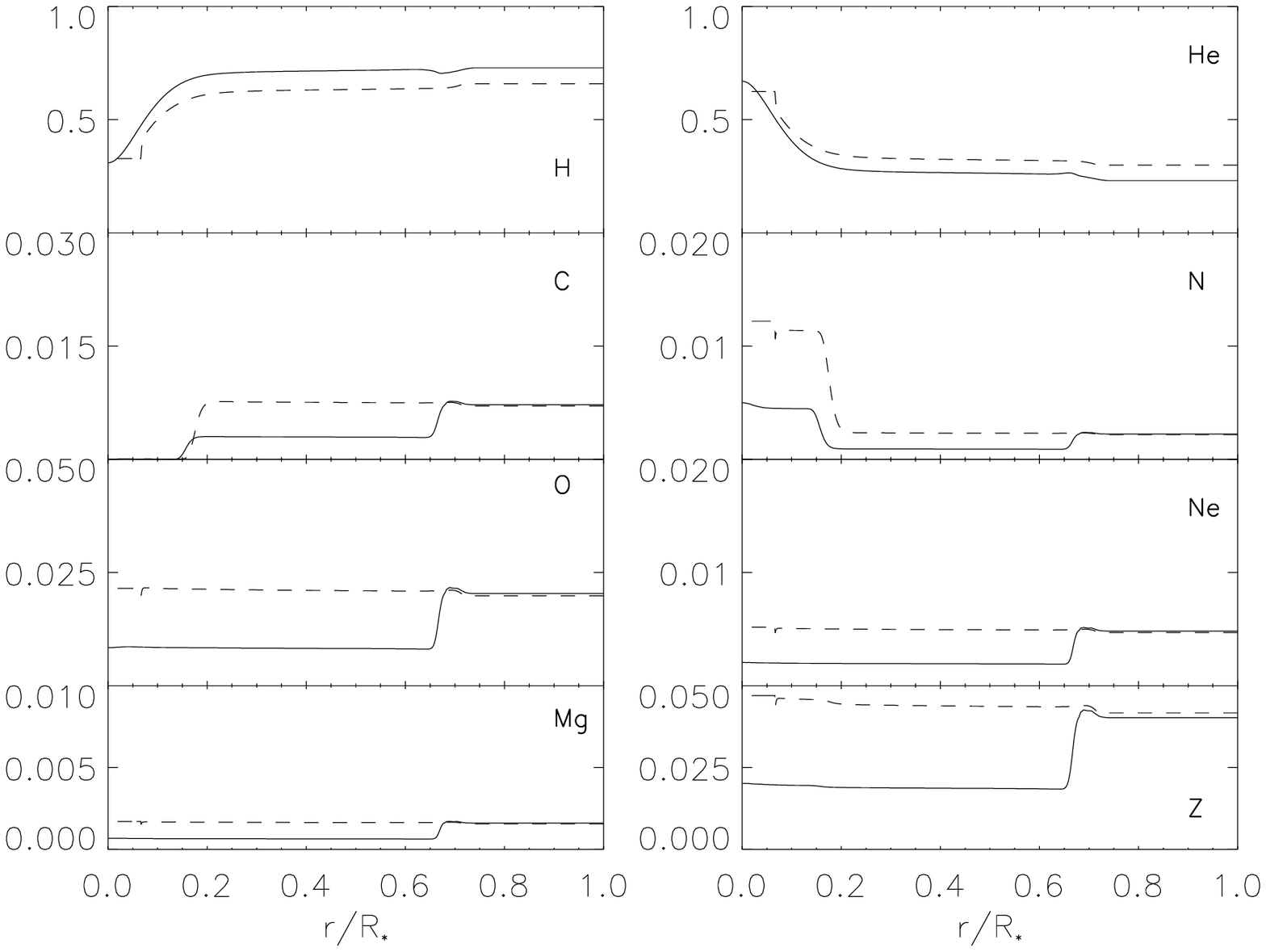}
\end{center}
\label{fig8}
\caption{Same as Fig.7 for models AC2 and OM2.}
\end{figure}

\subsection{Pulsation diagram} 

Figure 9 presents the pulsation diagram for each model. The Brunt-V\"ais\"al\"a frequency $N$ 
is plotted together with the Lamb frequency S$_{l}$ for $l = 1, 2, 3$ , as a function of 
the fractional radius. They were computed using the classical formulae :

\begin{equation}
N^2=\frac{g}{H_{p}} \left( \nabla_{ad}-\nabla_{rad}+ \nabla_{\mu} \right)
\end{equation}
\begin{equation}
S^{2}_{l}=l(l+1)\frac{c^2}{r^2}
\end{equation}

Horizontal lines are schematically drawn in Fig. 9, for frequencies of 1 mHz and 2.5 mHz, 
down to the curves $l=1$, to help visualize the resonant acoustic cavities.
The negative values of N$^2$ correspond to the region in which the medium is unstable against convection. 

\begin{figure}
\begin{center}
\includegraphics[angle=0,totalheight=\columnwidth,width=\columnwidth]{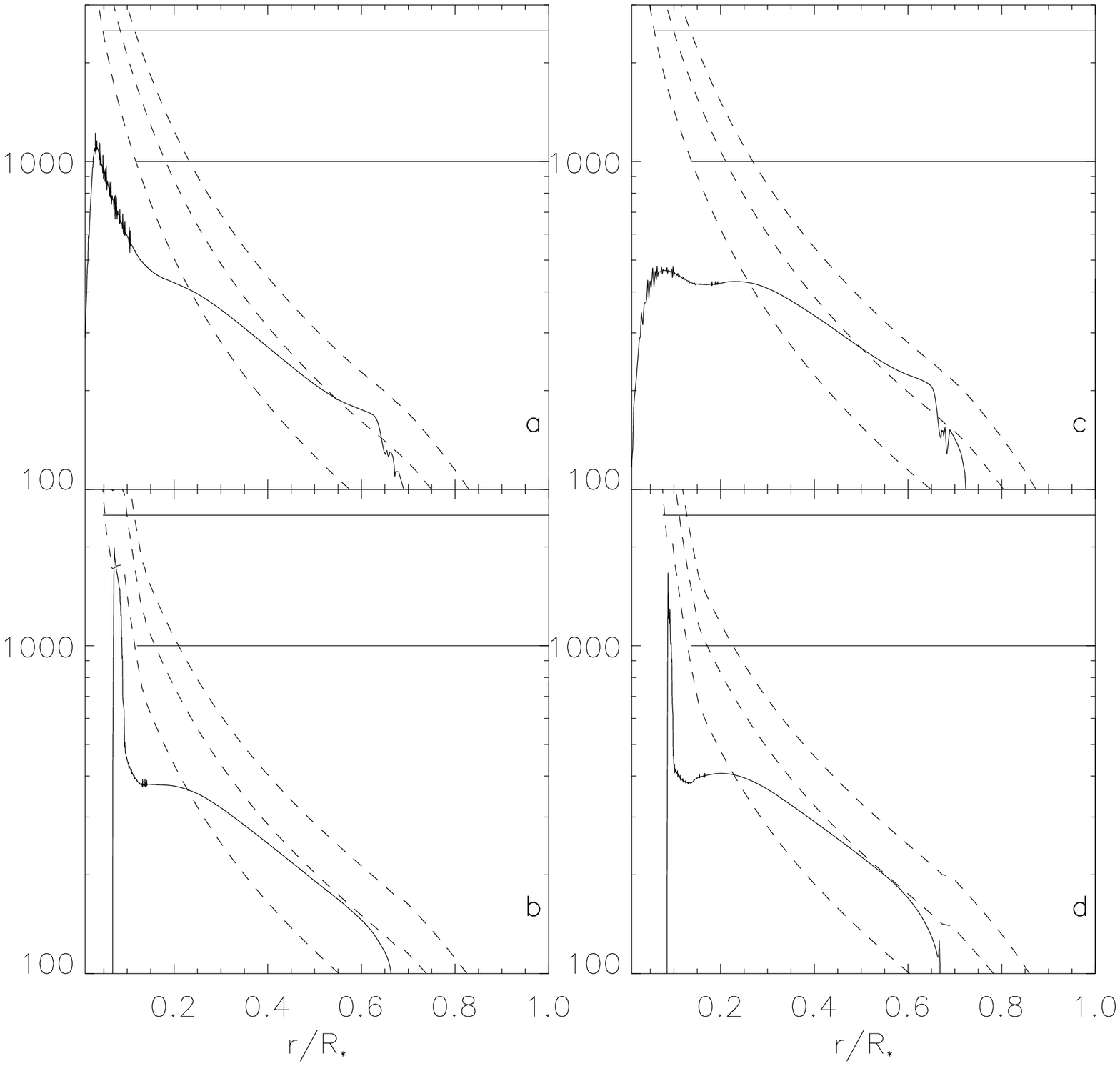}
\end{center}
\label{fig9}
\caption{Pulsation diagrams for AC1 ({\bf a}), OM1 ({\bf b.}), AC2 ({\bf c.}), OM2 ({\bf d.}).
 The full lines represent the Brunt-V\"ais\"al\"a frequency, the dashed lines represent 
the Lamb frequency for $l=1, 2, 3$. Horizontal lines are drawn in each graphs for frequencies
 of 1 mHz and 2.5 mHz to help visualize the acoustic cavities.}
\end{figure}

\section{Asteroseismic tests} 

For each model the oscillation frequencies were computed for p-modes of degrees $l = 0$ to
 $ l = 3 $. We then studied in detail the different combinations of 
these frequencies which could lead to observational tests of the internal structure and chemical
 composition of the stars.

\subsection{Large separations}

The ``large separations" represent the frequency differences between two modes of the same
 degree $l$ and successive radial numbers $n$ :

\begin{equation}
\Delta\nu_{n,l}=\nu_{n,l} - \nu_{n-1,l}
\end{equation}

For p-modes, the large separations are nearly constant, with values close to the fundamental 
asteroseismic quantity $\nu_0 = 1/2 t_{\star}$, where $t_{\star}$ is the stellar acoustic radius 
(Table 2) (e.g. Audard \& Provost \cite{audard94}).

Figures 10 and 11 display the ratio of the large separations for the four models, divided by $\nu_0$. 

\begin{figure}
\begin{center}
\includegraphics[angle=0,totalheight=\columnwidth,width=\columnwidth]{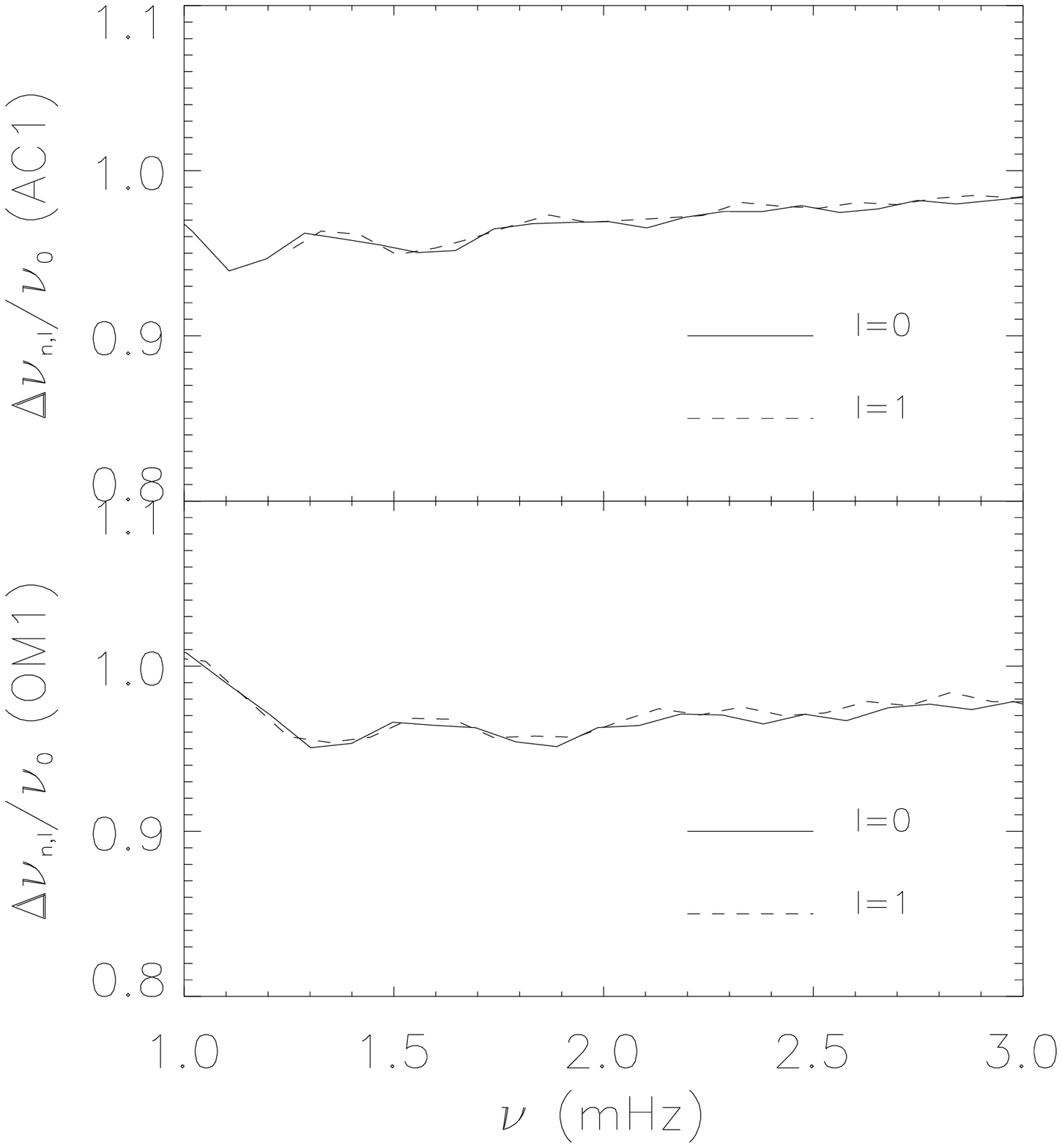}
\end{center}
\label{fig10}
\bigskip
\bigskip
\caption{Large separations devided by $\nu_0$, inverse of twice the acoustic radius of the star, for models OM1 and AC1.}
\end{figure}

\begin{figure}
\begin{center}
\includegraphics[angle=0,totalheight=\columnwidth,width=\columnwidth]{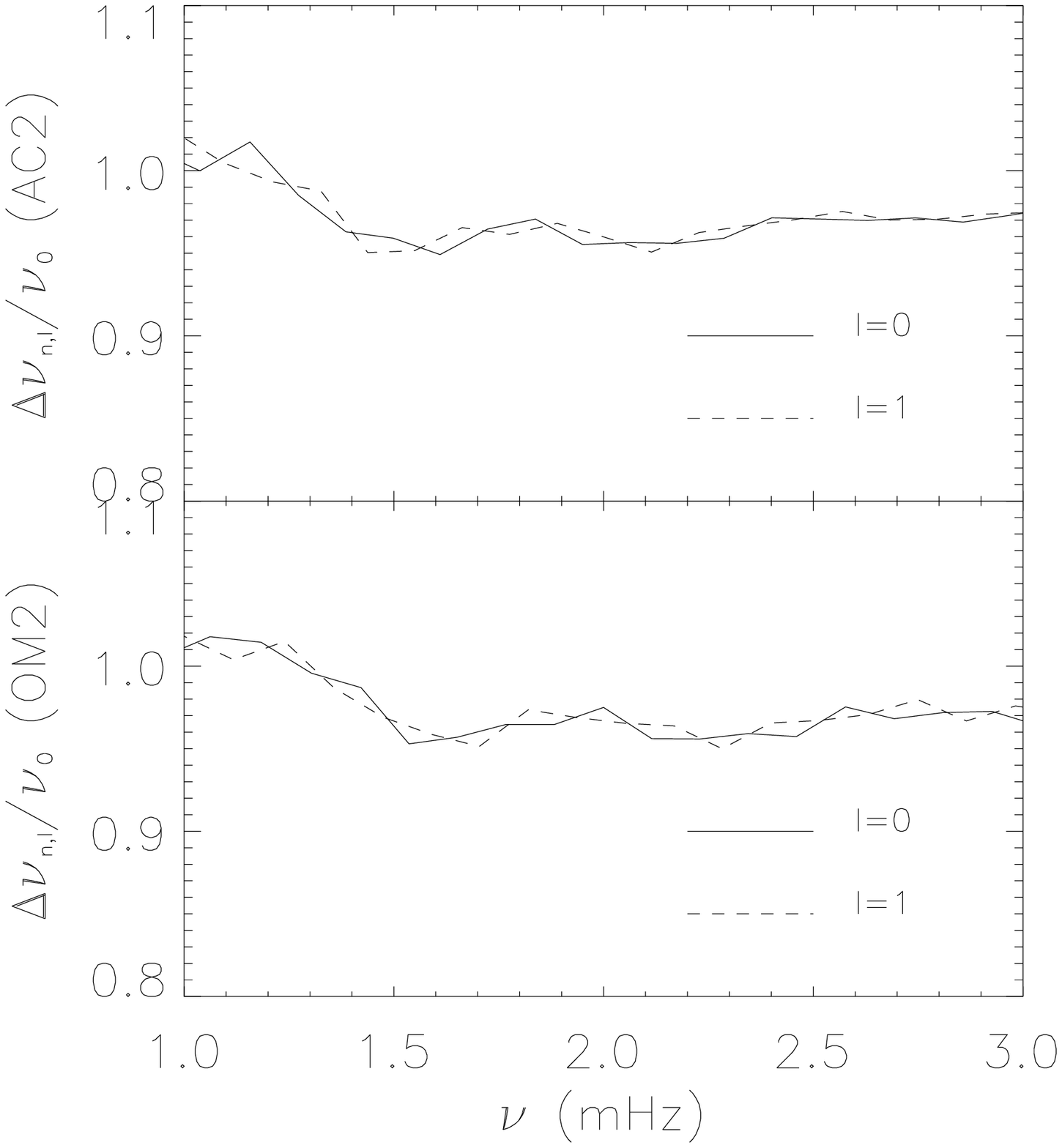}
\end{center}
\label{fig11}
\bigskip
\bigskip
\caption{Large separations devided by $\nu_0$, inverse of twice the acoustic radius of the star, for models OM2 and AC2.}
\end{figure}

\subsection{Small separations}

We define the ``small separations" as Roxburgh \& Voronstsov \cite{roxburgh01}:

\begin{equation}
D_{n,l}=\frac{\nu_{n,l}-\nu_{n-1,l+2}}{2l+3}
\end{equation}

The repercussion of the presence of a convective core on the small separations have been 
studied by several authors (Provost et al \cite{provost93}, Audard, Provost and
 Christensen-Dalsgaard \cite{audard95}, Roxburgh \& Voronstsov \cite{roxburgh01}).
 The boundary of the convective core induces a sharp variation in the sound velocity
 which leads to a partial reflexion of the waves which would otherwise travel down to
 the stellar center. Roxburgh \& Vorontsov \cite{roxburgh01} have computed the oscillatory
 perturbations which appear in this case in the small separations. They exhibit a modulation
 with a period close to the acoustic radius of the core (time needed for the waves
 to travel between the stellar center and the core boundary) (Table 2). The oscillations have
 opposite phases for $l = 0$ and $l = 1$. When the effect of gravity perturbations is taken
 into account, this acoustic radius has to be replaced by an ``effective acoustic radius" given by :

\begin{equation}
 \tilde{t} = \int_{t_{1}}^{t}\left( 1 + \frac{G \rho_{0}}{\pi \nu^{2}} 
\right)^{\frac{1}{2}} dt = \int_{r_{1}}^{r}\left( 1 + \frac{G 
\rho_{0}}{\pi \nu^{2}} \right)^{\frac{1}{2}} \frac{dr}{c}
\end{equation}

Roxburgh \& Vorontsov \cite{roxburgh01} have shown that the best parameter to test
 the presence of a convective core is indeed the ratio of the small to the large separations,
 namely $R_{n,l} = \frac{\pi}{2}\frac{D_{n,l}}{\Delta \nu_{n,l}}$, as it helps getting rid
 of the surface effects on the frequencies. 
Unfortunately, the characteristic oscillations ``periods" (in frequency units) are larger 
than the atmospheric cut-off frequency, so that only part of one oscillation can be visible
 from observations.

Figures 12 and 13 display the small separation for the four models, for $l = 0$ and $l = 1$ , 
up to frequencies of 10 mHz. In real stars the oscillation modes are only seen up to the cut-off 
frequency $\nu_{c} = \frac {c}{4 \pi H_p}$ (Mazumdar \& Antia \cite{mazumdar01}) which is indicated 
in the figures by a vertical
 line. The modulation is clearly seen in the overmetallic models, as well as the opposite
 phase between the modes of different degrees.
Figures 14 and 15 present, for the four models, the parameter $R_{n,l}$ as a function of the frequencies, 
limited to 3 mHz which is a value close to the cut-off frequencies. This figure gives a clear idea
 of the values which may be obtained from the observations of stellar oscillations.
 We can see that, although only part of a modulation period is displayed, the oscillatory
 character of the ratio $R_{n,l}$ appears for the overmetallic models but not for the accretion models. 

\begin{figure}
\begin{center}
\includegraphics[angle=0,totalheight=\columnwidth,width=\columnwidth]{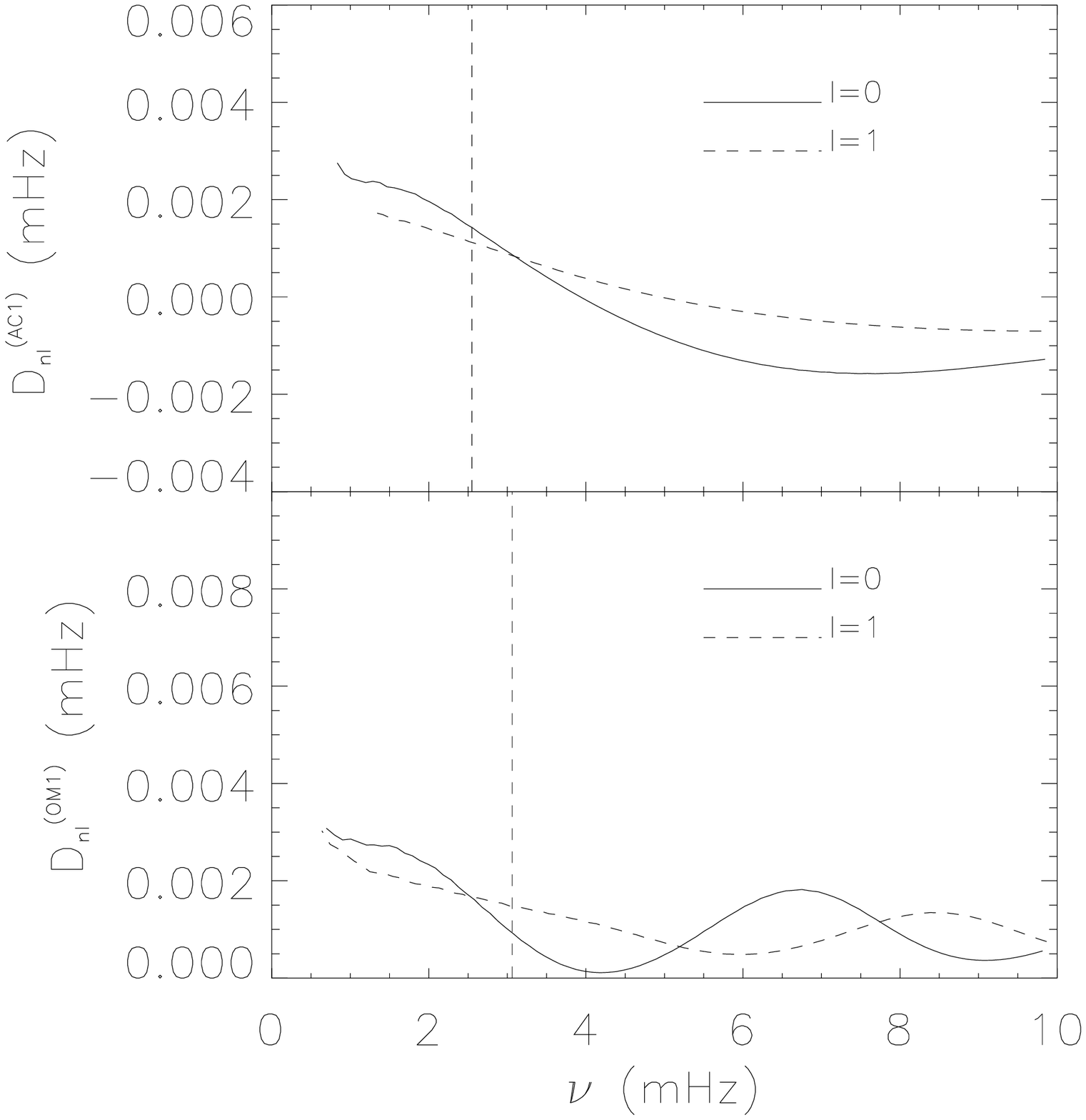}
\end{center}
\label{fig12}
\caption{ Small separation for models AC1 and OM1, for $l = 0$ and $l = 1$ , up to frequencies of 10 mHz.
The vertical line corresponds to the cut-off frequencies. The oscillations due to the convective core are
 clearly seen for the overmetallic model. However they cannot be as clearly visible in real stars due
 to the cut-off frequencies.}
\end{figure}

\begin{figure}
\begin{center}
\includegraphics[angle=0,totalheight=\columnwidth,width=\columnwidth]{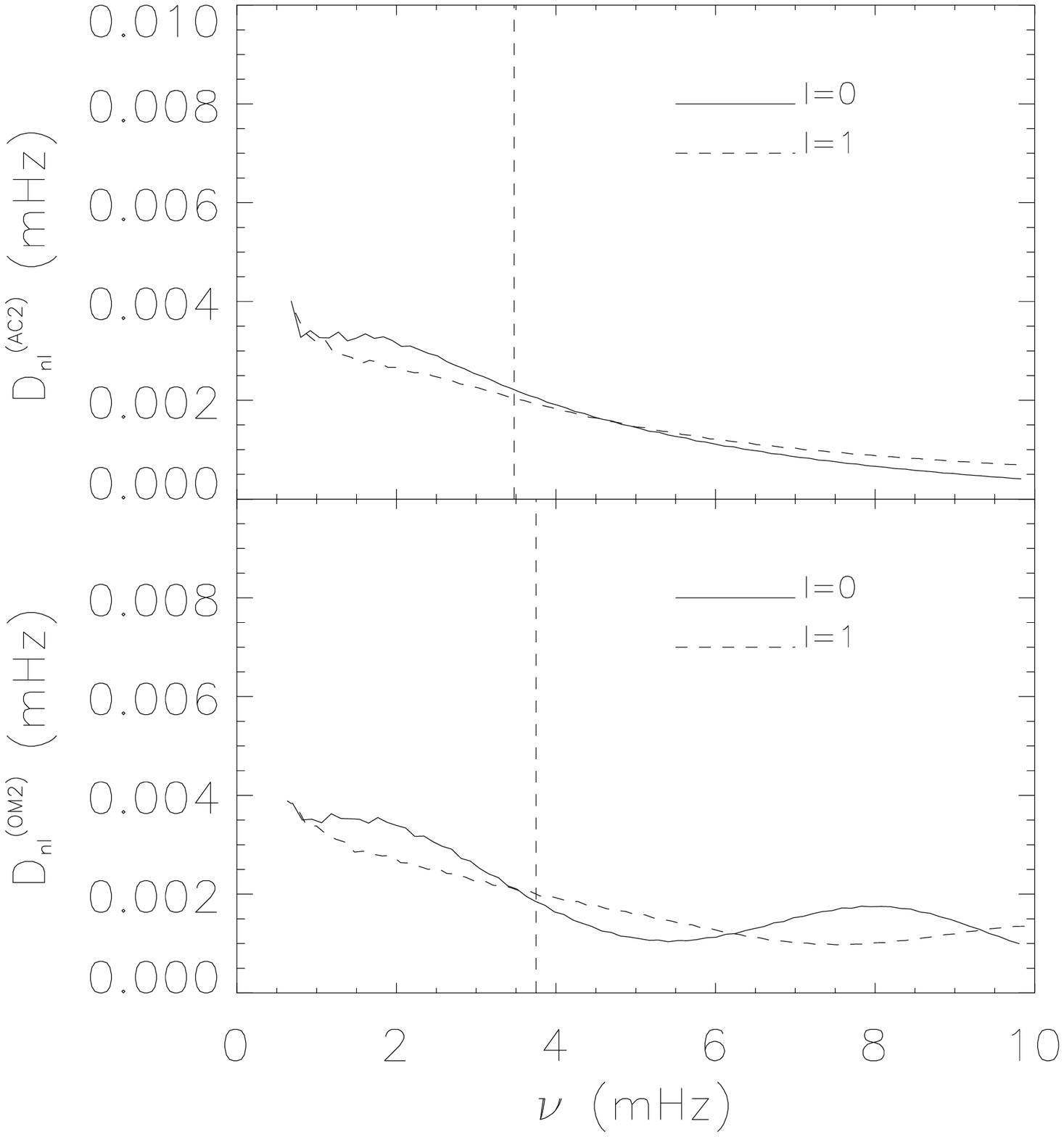}
\end{center}
\label{fig13}
\caption{ Same as Fig.12 for models AC2 and OM2.}
\end{figure}

\begin{figure}
\begin{center}
\includegraphics[angle=0,totalheight=\columnwidth,width=\columnwidth]{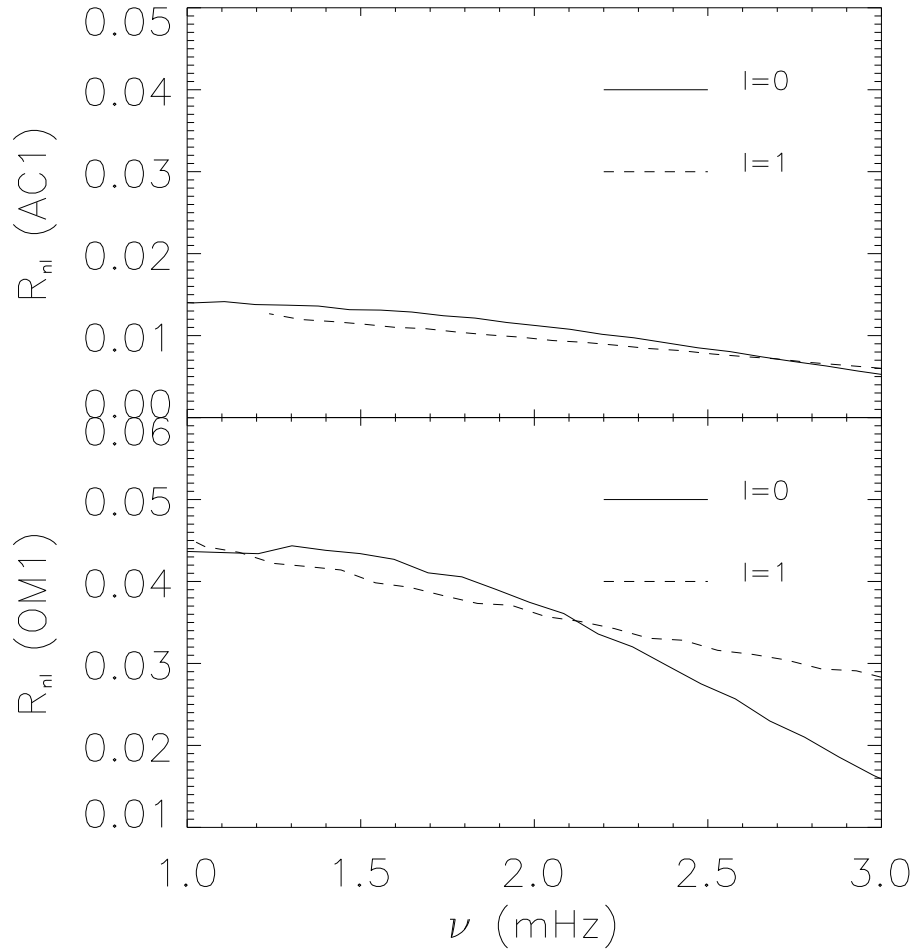}
\end{center}
\label{fig14}
\caption{ Parameter $R_{n,l} = \frac{\pi}{2}\frac{D_{n,l}}{\Delta \nu_{n,l}}$ , for models AC1 and OM1, up to
$\nu$ = 3 mHz, close to the cut-off frequencies; The beginning of the oscillations is evident in the
 overmetallic model for $l = 0$. }
\end{figure}

\begin{figure}
\begin{center}
\includegraphics[angle=0,totalheight=\columnwidth,width=\columnwidth]{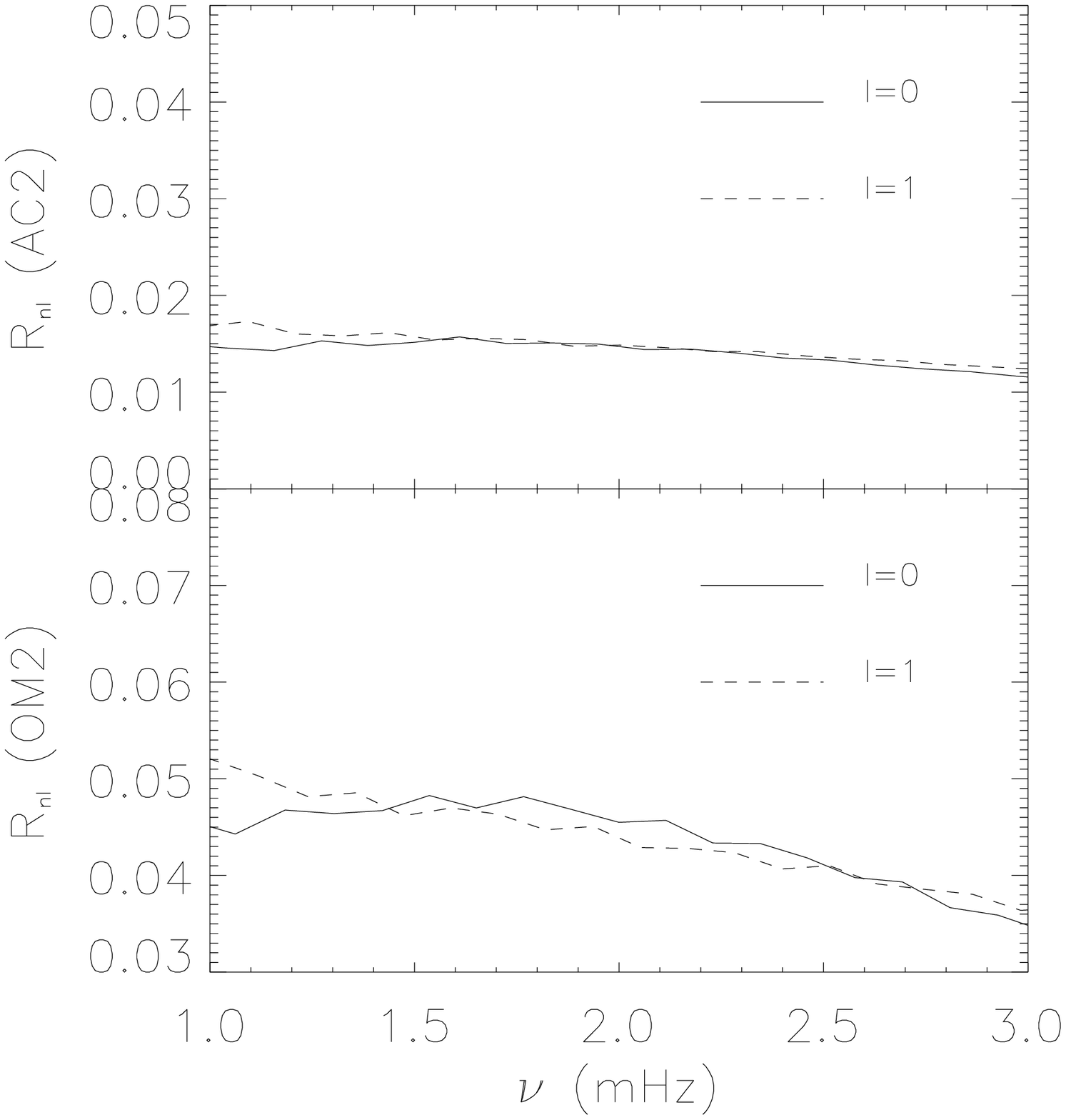}
\end{center}
\label{fig15}
\caption{ Same as Fig.14, for models AC2 and OM2}
\end{figure}

\subsection{Second differences}

Apart from their cores, overmetallic and accretion models also show some important differences 
in their outer layers. The most important is the presence of a steep $\mu$-gradient just below
 the convective zone of the accretion models. Also the sound velocities in the convective zones
 are larger by 2 to 5 \% for the OM models compared to the AC models.
Gough (\cite{gough90}) showed that rapid variations in the sound velocities in the outer layers
 of stars lead to oscillations which could be better seen in the so-called ``second differences"
 $\delta_2 \nu = \nu_{n+1} -2\nu_{n} + \nu_{n-1}$.

In the absence of chemical gradients, second differences are good indicators of the hydrogen and 
helium ionization zones, and of the base of the convective zone (Monteiro et al. \cite {monteiro94}, 
Monteiro \& Thompson \cite{monteiro98}, Miglio et al \cite{miglio03}). When a helium gradient is present
 below the convective zone, it becomes the most effective ``partial mirror" for the waves, 
leading to the most important modulation (Vauclair and Th\'eado \cite{vauclair04b}). 
On the other hand, the second differences are not influenced by the presence of a convective core 
(Mazumdar \& Antia \cite{mazumdar01}).

Figures 16 and 17 present the second differences for the four models, as well as their Fourier transforms, 
while Figure 18 displays the usual thermodynamical parameter $\Gamma_1$ as a function of fractional radius.
 Three peaks are observed in the Fourier transforms. The two ones on the left, below times of 2000 s,
 correspond to the ionisation zones which lead to the drops and kinks in $\Gamma_1$.
 Each peak appears for a characteristic time which is twice the time needed for the sound waves
 to travel from the surface down to the considered layers. The third peak, around 6000 s, 
corresponds to the bottom of the convective zone, as can be checked from the acoustic depth given 
in Table 2. In each case, the acoustic depth of the convective zone, hence the position of the peak, 
corresponds to a larger time in the accretion models than in the overmetallic models. Here the helium 
gradients are not strong enough, nor the metallic gradients induced by accretion, to be seen in the
 second differences.

\begin{figure}
\begin{center}
\includegraphics[angle=0,totalheight=\columnwidth,width=\columnwidth]{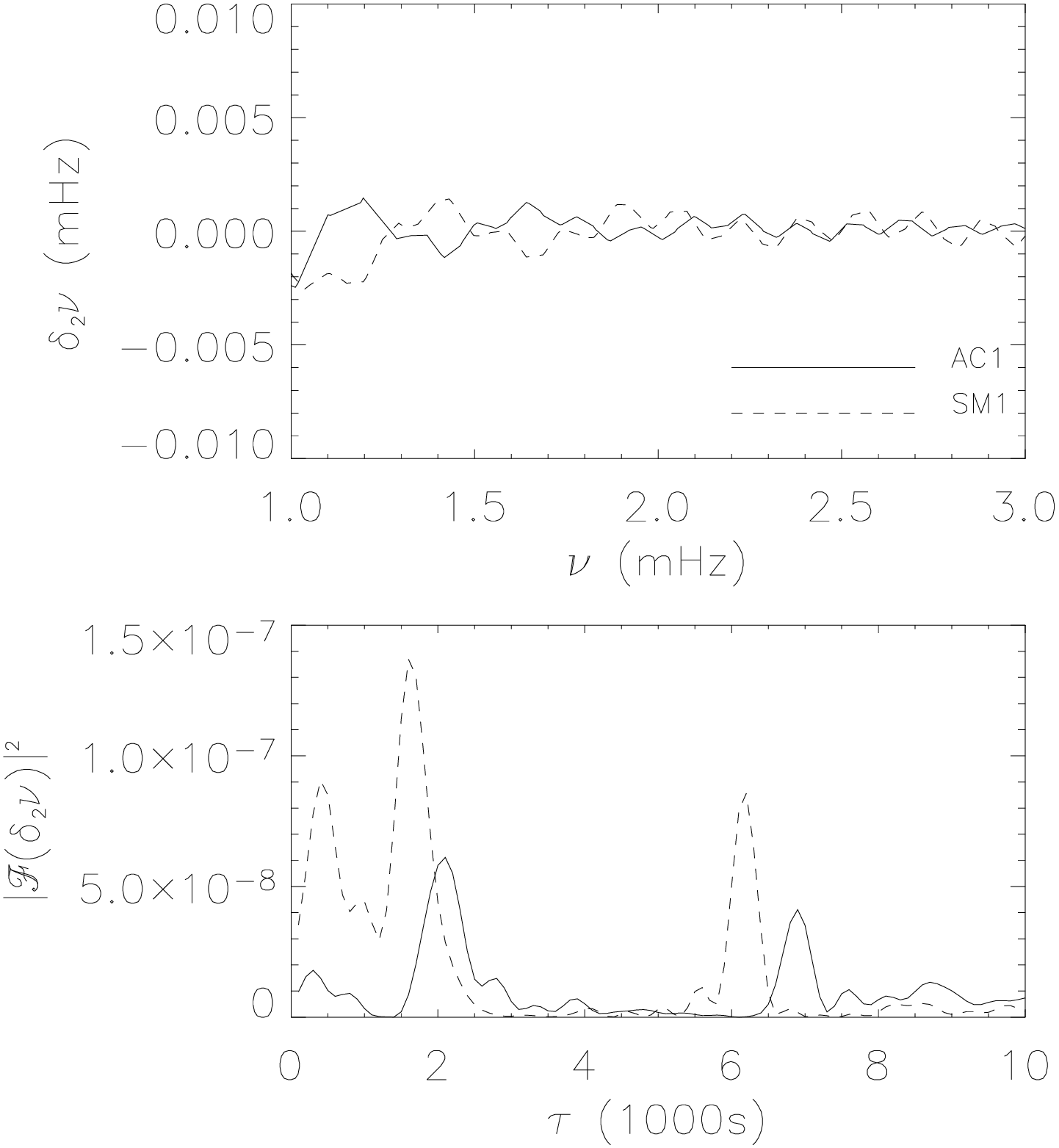}
\end{center}
\label{fig16}
\caption{ Second differences and Fourier transforms for models OM1 and AC1. The peaks below 2000 s are due to the helium
ionisation zones while the peak around 6000 s is due to the bottom of the convective zone.}
\end{figure}

\begin{figure}
\begin{center}
\includegraphics[angle=0,totalheight=\columnwidth,width=\columnwidth]{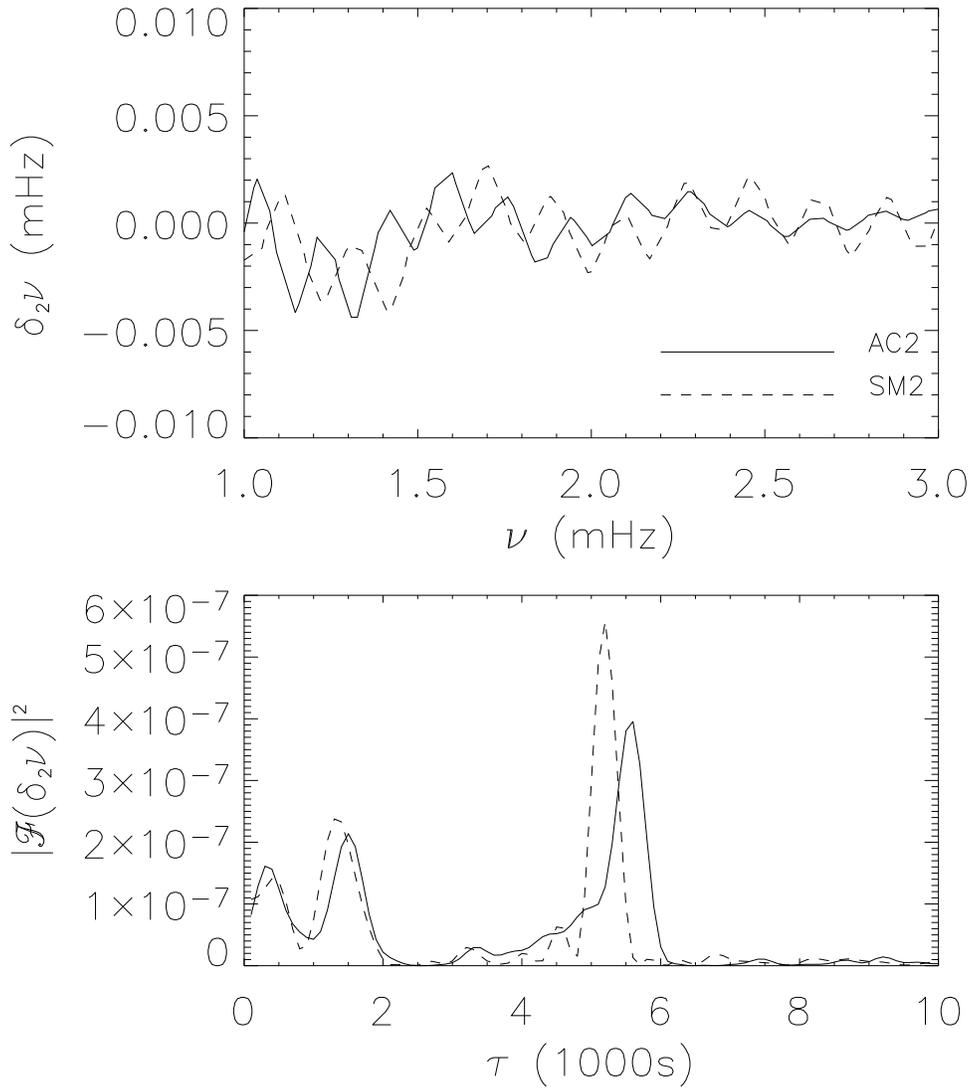}
\end{center}
\label{fig17}
\caption{ Same as Fig. 16 for models OM2 and AC2}
\end{figure}

\begin{figure}
\begin{center}
\includegraphics[angle=0,totalheight=\columnwidth,width=\columnwidth]{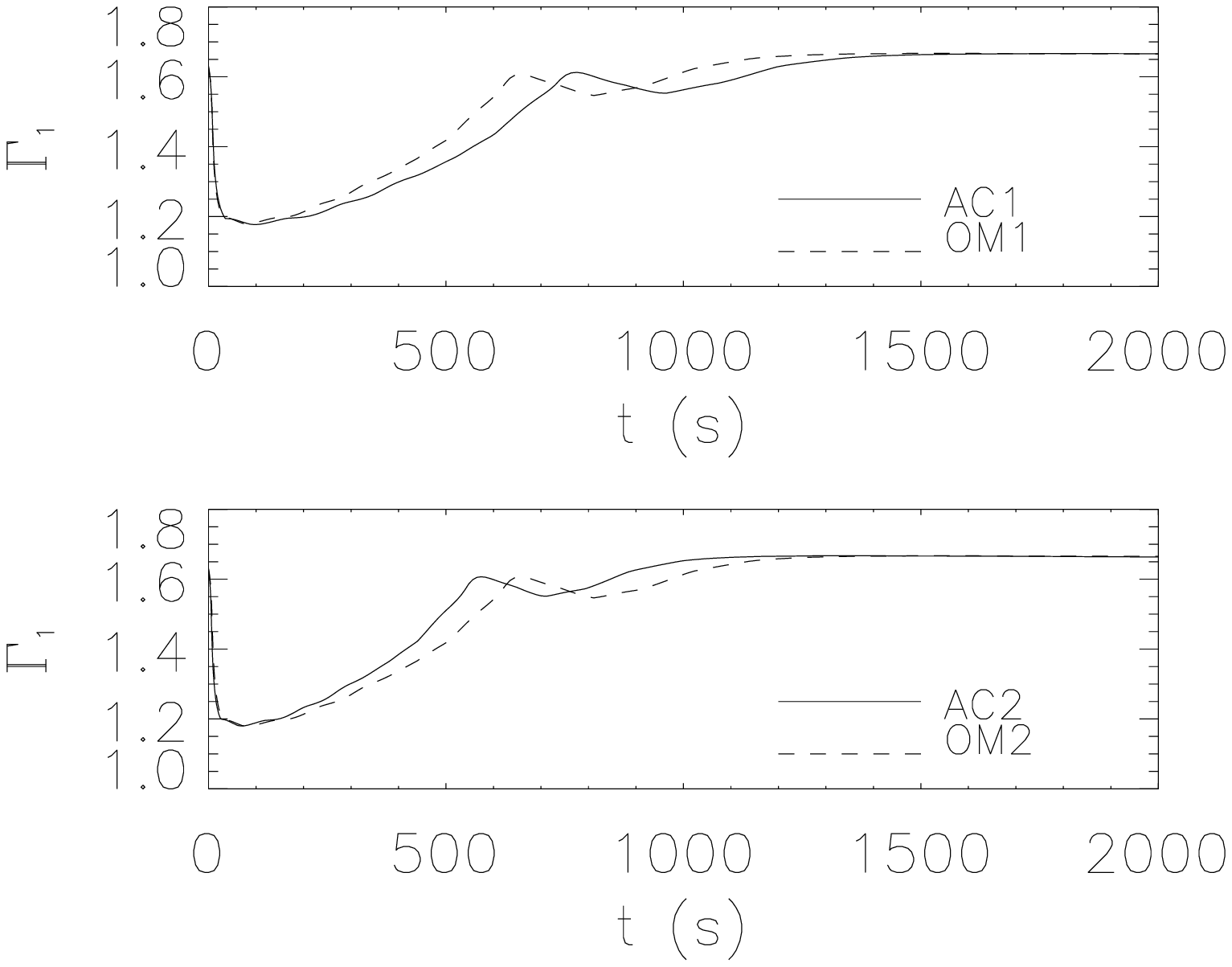}
\end{center}
\label{fig18}
\caption{ Profiles of the thermodynamical parameter $\Gamma_1$ for models OM1 and AC1 (top), OM2 and AC2 (bottom).
The kinks in $\Gamma_1$ are the reason for the two left peaks in Figs. 16 and 17, bottom graphs.}
\end{figure}

\section{Discussion}

We have studied two couples of models, each of them with identical outer (observable) 
parameters : $L, T_{eff}$, external chemical composition, but different histories. In each couple,
 one partner was a model with an overmetallicity down to the stellar center (overmetallic model)
 while the other partner was a model with normal (solar) composition with an metallicity increase 
in the external layers due to accretion (accretion model). In the first couple (models OM1 and AC1),
 the helium value was solar in the two partners, while in the second couple (models OM2 and AC2),
 the overmetallic model was assumed to have also an increase of helium, according to the Y-Z 
relation obtained for the chemical evomution of galaxies (Izotov \& Thuan \cite{isotov04}).

We always considered that the accreted matter contained neither hydrogen nor helium. 
For the other elements, we assumed solar relative abundances. The accreted material 
was supposed to be instantaneously mixed in the convective zone at the beginning
 of the zero-age-main-sequence. For a first approach, double-diffusive mixing
 (Vauclair \cite{vauclair04}) was not introduced in the present computations. Models including this effect will be presented in a forthcoming paper.

We show that their are drastic differences between two models which would appear as
 the same star from outside, but which have different chemical composition inside.
 The overmetallic models have a larger mass than the accretion models, and the most 
important differences in their internal structure is that the first ones  develop a
 convective core while the second ones do not. Also the outer convective zones do not 
have the same depth and there is a metallic gradient below the convective boundary in 
the accretion models.

We have investigated the possible asteroseismic tests of these differences. We found 
that the differences in the internal structure lead to signatures in the large separations
 (the acoustic radii of the stars are different), small separations and second differences.
 While the second differences can give the position of the bottom of the outer convective zone, 
the most important effect is seen in the ratio of the small to the large separations : 
an oscillatory behavior clearly related to the presence of the convective core is seen 
in the overmetallic models and does not exist in the accretion models.

These first results are very encouraging and we may hope to be able to derive, by observing
 real central stars of planetary systems, whether they are overmetallic down to the center or
 suffered planetary accretion. Such results would be very helpful to understand the formation 
of planetary systems.

\begin{acknowledgements}

The authors thank S. Charpinet for providing his pulsation code and for stimulating discussions.
Sylvie Vauclair acknowledges a grant from Institut universitaire
de France.

\end{acknowledgements}

\end{document}